\pdfoutput=1

\documentclass[12pt, a4paper]{iopart}
\usepackage{graphicx,epsf,hyperref}
\eqnobysec

\def\be{\begin{equation}}
\def\ee{\end{equation}}
\def\bea{\begin{eqnarray}}
\def\eea{\end{eqnarray}}
\def\vp{{\varphi}}
\newcommand\dvp[1]{{\delta\varphi_{#1}}}
\newcommand\dvpI[1]{{\delta\varphi_{{#1}I}}}
\newcommand\dvpK[1]{{\delta\varphi_{{#1}K}}}
\newcommand\dvpM[1]{{\delta\varphi_{{#1}M}}}
\newcommand\dvpN[1]{{\delta\varphi_{{#1}N}}}
\newcommand\dvpL[1]{{\delta\varphi_{{#1}L}}}

\newcommand\dU[1]{{\delta U_{#1}}}
\def\H{{\cal H}}
\def\cs2{c_{\rm{s}}^2}
\def\U0{{\bar U_0}}
\def\12{\frac{1}{2}}
\def\BkBk{{B_{1,k}B_{1,}^{~k}}}

\def\dvpdvpKll{\delta\vp_{1K,l}\delta\vp_{1K,}^{~~~~l}}

\newcommand\eq[1]{Eq.~(\ref{#1})}
\newcommand\eqs[1]{Eqs.~(\ref{#1})}

\begin{document}

\title{A short note on the curvature perturbation at second order}

\author{Adam J.~Christopherson$^{1,2,*}$, Ellie Nalson$^3$ and Karim A.~Malik$^3$}

\address{$^1$Department of Physics, University of Florida, Gainesville, FL 32611, USA}
\address{$^2$School of Physics and Astronomy, University of Nottingham, University Park,
Nottingham, NG7 2RD, UK}
\address{$^3$Astronomy Unit, School of Physics and Astronomy, Queen Mary University of London,
Mile End Road, London, E1 4NS, UK}

\ead{$^*$achristopherson@gmail.com}

\date{\today}

\begin{abstract}
Working with perturbations about an FLRW spacetime, we compute the
gauge-invariant curvature perturbation to second order solely in terms of scalar field fluctuations. Using the curvature perturbation on uniform density hypersurfaces as our starting point, we give our results in terms of field fluctuations in the flat gauge, incorporating both large and small scale behaviour. For ease of future numerical
implementation we give our result in terms of the scalar field
fluctuations and their time derivatives. 
\end{abstract}

\maketitle

\section{Introduction}

The first detection of the anisotropies in the cosmic microwave
background by the COBE satellite \cite{COBE} paved the way for the
predictions of the standard cosmological model to be tested. In the
decades since, experiments have tremendously increased in their
sophistication enabling us to put ever stronger constraints on our
theory of the early universe \cite{WMAP7, Ade:2013uln}. The most recent experimental success was the detection of B-mode polarization by the BICEP experiment \cite{Ade:2014xna}. Although a large fraction of this signal is now explained by galactic foregrounds \cite{Ade:2015tva}, there is still room for a signal of primordial gravitational waves from inflation. This would be a huge success for inflationary cosmology.

The initial inhomogeneities in the density field are generated by
quantum fluctuations during inflation. In the simplest models this is
a single field slowly rolling down a potential, but there are a
plethora of models with multiple fields, fields with complicated
kinetic terms, non-minimally coupled fields,
etc.~(see e.g.~Ref.~\cite{Baumann:2009ds}).
In order to model the inhomogeneities in the universe we use
cosmological perturbation theory. This involves taking the homogeneous
Friedmann-Lema\^{i}tre-Robertson-Walker (FLRW) solution as a background
spacetime and adding small, inhomogeneous perturbations on top
(e.g. \cite{Bardeen:1980kt, ks, mfb}). The linear theory is a huge
success, matching the CMB and large scale structure excellently,
however given the quality of the data that we are now fortunate to
obtain, we can use higher order theory to put more restrictive
constraints on models, and aim to rule out large classes of models.

In this article we focus on one particular quantity: the curvature
perturbation on uniform density hypersurfaces, $\zeta$. This gauge
invariant quantity is the variable of choice when computing, say, the
bispectrum of inflationary fluctuations. We derive the expression
giving the curvature perturbation at second order in perturbation
theory, $\zeta_2$, in terms of scalar field variables, allowing for
multiple minimally coupled scalar fields. Previous studies have either focused on large scales or have included a limited number of fields, either using perturbation theory or the $\delta N$ formalism (for a non-exhaustive list of references see Refs.~\cite{Starobinsky:1986fxa, Sasaki:1995aw, Sasaki:1998ug, Wands2000, Rigopoulos:2003ak, Lyth:2004gb, Malik2004, Lyth:2005du, Malik:2005cy, Langlois:2005qp, Nakamura:2010yg, Tzavara:2011hn}). We envisage our calculation
to be of particular use for future numerical computations of
bispectra or other inflationary parameters and, to that end, ensure
that the resultant expression is in a closed form, containing only
field fluctuations and their single time derivative.

The paper is structured as follows: in the next section we present the
governing equations for perturbations up to second order. In
Section \ref{sec:zeta} we define the curvature perturbation, and
present our main result. Then, in Section \ref{sec:results}, we specialise to the case of a single field in the slow-roll limit. We close with a discussion in
Section \ref{sec:diss}.  We assume an FLRW background spacetime with
zero spatial curvature, and use conformal time, $\eta$,
throughout. Greek indices, $\mu,\nu,\lambda$, cover the entire
spacetime range, from $0,\ldots 3$, while we use lower case Latin,
$i,j,k$, to denote spatial indices running from $1,\ldots3$. Upper
case Latin indices, $I,J,K$, denote different scalar fields.

\section{Governing equations}
\label{governing_sect}

In this section we review the governing equations for the cosmological
perturbations up to second order that we will use when computing the
curvature perturbation for a system containing multiple scalar
fields. We follow the notation and definitions of, e.g.,
Refs.~\cite{MW2008, thesis} throughout. In this paper we consider on scalar perturbations to FLRW, for which the line element takes the form 
\be 
ds^2=a^2(\eta)\Bigg[-(1+2\phi)d\eta^2+2aB_{,i}dx^id\eta
+\Big((1-2\psi)\delta_{ij}+2E_{,ij}\Big)dx^idx^i\Bigg]\,,
\ee
where perturbed quantities are then expanded, order by order, as, e.g., $\phi=\phi_1+\frac{1}{2}\phi_2$. In the uniform curvature gauge, $E=\psi=0$, and the two scalar degrees of freedom in the metric are $\phi$ and $B$.

\subsection{The field equations}
\label{field_sect}

The governing field equations are obtained by perturbing the Einstein equation
$
G_{\mu\nu}=8\pi GT_{\mu\nu}\,,
$
and truncating at the required order. In the background, we obtain the usual Friedmann and acceleration equations:
\be
\label{Friedmann}
\H^2=\frac{8\pi G}{3}  
\left(\sum_K\frac{1}{2}{\vp'_{0K}}^2+a^2 U_0\right)  \,,
\ee
%
%
\be
\label{ij_back}
\left(\frac{a'}{a}\right)^2-2\frac{a''}{a}=8\pi G 
\left(\sum_K\frac{1}{2}{\vp'_{0K}}^2-a^2 U_0\right)  \,.
\ee
In the following, we will now present the linear and second order field equations in the uniform curvature gauge.

\subsubsection{First order}
\label{sec:field1}

The $0-0$ component of the Einstein equations at first order is
\be
\label{00Ein1}
2a^2U_0{\phi_1}+\sum_K\vp_{0K}'{{\dvpK1}}'+a^2{\delta U_1}
+\frac{\H}{4\pi G}\nabla^2B_1
=0\,,
\ee
where we have used the background equations from above to simply the
expression. The $0-i$ part gives
\be
\label{0iEin1}
\H\phi_1-4\pi G\sum_K\vp_{0K}'{{\dvpK1}}=0\,.
\ee
Although they are not needed in the following, for completeness we
also give, from the spatial component of the Einstein equations, the
trace free equation
\be
\label{offtrace1}
B_1'+2\H B_1+\phi_1=0\,,
\ee
governing the evolution of the shear, and, using \eq{offtrace1}
and the background field equations, the first order trace in its
simplest form is
\be
\label{trace1}
\H\phi_1'+4\pi G\left[ 
a^2\delta U_1+2a^2U_0{\phi_1}-\sum_K\vp_{0K}'{{\dvpK1}}'\right]
=0\,.
\ee
\eq{00Ein1} and \eq{0iEin1} above can then be combined to give
\be
\nabla^2B_1=-\frac{4\pi G}{\H}\sum_K\Big(X_K\dvpK1
+\vp_{0K}'{{\dvpK1}}' \Big)\,,
\ee
where, following Ref.~\cite{Malik:2006ir}, we have defined
\be
\label{defXI}
X_K\equiv a^2\left(
\frac{8\pi G}{\H}U_0\vp_{0K}'+U_{,\vp_K} \right)\,.
\ee
%

\subsubsection{Second order}
\label{sec:field2}

Using the first order and background $0-0$ equations, \eqs{Friedmann} and (\ref{00Ein1}), the $0-0$ component of the second order Einstein equation is
\bea
\label{Ein00_2}
\fl
8\pi G a^2 U_0\left(\phi_2+\BkBk\right)
+\H\nabla^2B_2 
+\12\left[B_{1,kl}B_{1,}^{~kl}
-\left(\nabla^2 B_1\right)^2\right]
-2\H \phi_{1,k}B_{1,}^{~k}\nonumber\\
\fl \qquad+4 \pi G\sum_K\left[
\vp_{0K}'\dvpK2' +a^2\delta U_2+4 a^2\delta U_1\phi_1
+{\dvpK1'}^2+\dvpK1_{,k}\dvpK1_{,}^{~k}
\right]=0\,,
\eea
while the $0-i$ Einstein equation is given by
\bea
\label{0i_2}
\H\phi_{2,i}-4\H\phi_{1}\phi_{1,i}+2\H B_{1,ki}B_{1,}^{~k}
+B_{1,ki}\phi_{1,}^{~k}-\nabla^2 B_1\phi_{1,i}\nonumber\\
\qquad\qquad
-4\pi G \sum_K\left[\vp_{0K}'\dvpK2_{,i}+2\dvpK1'\dvpK1_{,i}\right]=0\,.
\eea
We then use \eq{0iEin1}, rewrite
\eq{0i_2} and take the trace, to obtain
%
%
\bea
\fl
\label{0i_2version2}
\H\left(\phi_{2}-2\phi_{1}^2+ B_{1,k}B_{1,}^{~k}\right)
-4\pi G \sum_K\vp_{0K}'\dvpK2
+\nabla^{-2}\left(\phi_{1,kl}B_{1,}^{~~kl}
-\nabla^2 B_1\nabla^2\phi_1\right)\nonumber\\
\fl\qquad\qquad
-8\pi G \sum_K \nabla^{-2}\left(
\dvpK1'\nabla^{2}\dvpK1+\delta\vp_{1K,l}'\delta\vp_{1K,}^{~~~~l}
\right)
=0\,,
\eea
where we have introduced the inverse Laplacian, defined as
$\nabla^{-2}(\nabla^{2}X)=X$,
and have utilised the following identities
\bea
\nabla^2\left(\phi_1^2\right)
&=&2\left(\phi_1\nabla^2\phi+\phi_{1,k}\phi_{1,}^{~k}\right)\,, \nonumber \\
\nabla^{2}\left(B_{1,k}B_{1,}^{~k}\right)
&=&2\left(B_{1,k}\nabla^{2}B_{1,}^{~k}
+B_{1,kl}B_{1,}^{~kl}\right)\,.
\eea
%

\subsection{The Klein-Gordon equation}
\label{sec:KG}

The evolution of the scalar fields is given by the
Klein-Gordon equation, obtained from energy-momentum conservation. In
the background, this gives
\be
\label{KGback}
\vp_{0I}''+2\H\vp_{0I}'+a^2 U_{,\vp_I}=0\,,
\ee
 and for linear scalar field fluctuations is
 \be
 \fl
\label{KG1flatsingle}
\dvpI1''+2\H\dvpI1'+2a^2 U_{,\vp_I}\phi_1
-\nabla^2\dvpI1-\vp_{0I}'\nabla^2 B_1
-\vp_{0I}'\phi'_1+a^2 \sum_K U_{,\vp_I\vp_K}\dvpK1
=0\,.
\ee
The latter can then be written in a closed form, using the linear
Einstein equations (see e.g.~Ref.~\cite{Malik:2006ir}), as
\bea
\label{KG1_flat}
\fl
\dvpI1''+2\H\dvpI1'-\nabla^2\dvpI1\\
\fl\qquad\qquad
+a^2\sum_K\left\{
U_{,\vp_K\vp_I}
+\frac{8 \pi G}{\H}\left(
\vp_{0I}'U_{,\vp_K}+\vp_{0K}'U_{,\vp_I}
+\vp_{0K}'\vp_{0I}'\frac{8 \pi G}{\H}U_0
\right)
\right\}\dvpK1=0\,.
\nonumber
\eea
For completeness, we present the Klein-Gordon equation at second order in the Appendix.

\section{The curvature perturbation}
\label{sec:zeta}

In order to arrive at meaningful calculations within cosmological
perturbation theory, we have to consider gauge invariant
quantities. There are several gauge invariant curvature perturbations,
including the comoving curvature perturbation, $\mathcal{R}$
\cite{Lyth:1984gv}, the Newtonian curvature perturbation $\Psi$
\cite{Bardeen:1980kt,mfb} and the curvature perturbation on uniform density
hypersurfaces, $\zeta$, \cite{Bardeen:1983qw}. In this article we will focus on
the latter, and in this section will present our major result of the
paper and compute $\zeta$ in terms of scalar field fluctuations.

\subsection{The curvature perturbation, $\zeta$, in terms of the energy density}
\label{zeta_rho}

The curvature perturbation on uniform density hypersurfaces, $\zeta$
is given, at first order in terms of fluid variables as
\begin{equation}
\label{defzeta}
-\zeta_1 \equiv  \psi_1 + \H \frac{\delta\rho_1}{\rho_0'} \,,
\end{equation}
which, when the RHS is evaluated in the uniform curvature gauge,
simply gives
\be 
\zeta_1=-\H \frac{\delta\rho_1}{\rho_0'}\,.
\ee

At second order the expression is more complicated and, expressed in the uniform curvature gauge, is
\bea
\label{zeta2scal}
\fl
-\zeta_2=\frac{\H}{\rho_0'}\delta\rho_2
-2\H\frac{\delta\rho_1\delta\rho_1'   }{{\rho_0'}^2}
-\H^2\left(5+3\cs2\right)\left(\frac{\delta\rho_1}{\rho_0'}\right)^2
+\frac{1}{2}\frac{\delta\rho_{1,k}\delta\rho_{1,}^{~k} }{{\rho_0'}^2}
+\frac{1}{\rho_0'}B_{1,k}\delta\rho_{1,}^{~k}\nonumber\\
\fl\qquad-\frac{1}{2}\nabla^{-2}\left\{\left[
\frac{\delta\rho_{1,}^{~~i}\delta\rho_{1,}^{~~j} }{{\rho_0'}^2}
+\frac{2}{\rho_0'}\delta\rho_{1,}^{~(i}B_{1,}^{~~j)}
\right]_{,ij}\right\}\,.
\eea

Note that there are different definitions for $\zeta$. In this work we follow the review by Malik \& Wands \cite{MW2008}, which can be related to the definition of $\zeta$ introduced by Salopek \& Bond \cite{Salopek:1990jq} through
\be 
\zeta_{\rm MW}=\zeta_{\rm SB}+\zeta_{\rm SB}^2\,.
\ee

\subsection{Relating fluid to field variables}
\label{Relate}

In this section we present the relationship between the fluid
variables and scalar field variables. We simply quote the results
here, which are obtained by comparing the energy momentum tensor
components for a perfect fluid and a collection of minimally coupled
scalar fields, see Refs.~\cite{Malik:2006ir,MW2008}.

In the background the relationships are
\bea
\rho_0&=&\sum_K\frac{1}{2a^2}{\vp'_{0K}}^2+U_0\,, \\
P_0&=&\sum_K\frac{1}{2a^2}{\vp'_{0K}}^2-U_0\,.
\eea
Considering now perturbations in the uniform curvature gauge, we find
that the density perturbation at first order is in terms of the field
fluctuations
\be
\label{eq:drho1}
\delta\rho_1
=\frac{1}{a^2}\sum_K\left(\vp_{0K}'\dvpK1'-{\vp'_{0K}}^2\phi_1\right)
+\dU1   \,,
\ee
while the equivalent expression at second order is 
\bea
\label{eq:drho2}
\fl
\delta\rho_2=
\frac{1}{a^2}\sum_K\Big[
\vp_{0K}'\dvpK2'-4\vp_{0K}'\phi_1\dvpK1'-{\vp'_{0K}}^2\phi_2
+4{\vp'_{0K}}^2\phi_1^2+\dvpK1'^2+a^2\dU2 \nonumber\\
\fl\qquad\qquad\qquad\qquad
+\dvpdvpKll-{\vp_{0K}'}^2\BkBk \Big] \nonumber\\
\fl\qquad- \frac{2}{a^2}\sum_K\vp_{0K}'{\dvpK1}_{,l}\left[
\frac{\sum_L\vp_{0L}'{\dvpL1}_{,}^{~~l}}{\sum_M{\vp_{0M}'}^2}
-\frac{4\pi G}{\H}\nabla^{-2}\sum_L\Big(X_L\dvpL1
+\vp_{0L}'{{\dvpL1}}' \Big)_{,}^{~~l}
\right]
\,,
\eea
where we have defined
\bea
\label{eq:defdU1}
\dU1&=&\sum_K U_{,\vp_K}  \dvp{1K}\,,\\
\label{eq:defdU2}
\dU2&=&\sum_{K,L}U_{,\vp_K\vp_L}\dvp{1K}\dvp{1L}
+\sum_K U_{,\vp_K}\dvp{2K}\,.
\eea
The energy density perturbations can be written in terms of only
scalar field quantities by using the Einstein field equations at first
and second order. This gives
\be 
\delta \rho_1=\frac{1}{a^2}\sum_K\vp_{0K}'\Big(\dvpK1'-\vp_{0K}'\frac{4\pi G}{\H}\sum_L\vp_{0L}'\dvpL1\Big)+\delta U_1\,,
\ee
and
\bea 
\fl
\delta\rho_2=\frac{1}{a^2}\sum_{K,L,M}
\Bigg[\vp_{0K}'\dvpK2'+a^2\delta U_2+\dvpK1'^2
+\dvpK1_{,l}\dvpK1_{,}{}^l
-\frac{16\pi G}{\H}\vp_{0K}'\vp_{0L}'\dvpK1'\dvpL1
\nonumber\\
\fl\quad
-\frac{4\pi G}{\H}\vp_{0K}'^2\Bigg\{\vp_{0L}'\dvpL2
-\frac{8\pi G}{\H}\vp_{0L}'\vp_{0M}'\dvpL1\dvpM1
+2\nabla^{-2}\Big(\dvpL1'\nabla^2\dvpL1+\dvpL1_{,l}'\dvpL1_,{}^l\Big)
\nonumber\\
\fl\quad
-\frac{4\pi G}{\H^2}\nabla^{-2}\Big(\vp_{0M}'\nabla^2\dvpM1
(X_L\dvpL1+\vp_{0L}'\dvpL1')
-\vp_{0L}'\dvpL1_{,kl}\nabla^{-2}(X_M\dvpM1+\vp_{0M}'\dvpM1')_,{}^{kl}\Big)\Bigg\}\Bigg]
\nonumber\\
\fl\quad
-\frac{2}{a^2}
\sum_L\vp_{0L}'\dvpL1_{,l}\Bigg[\frac{\sum_K\vp_{0K}'\dvpK1_,{}^l}{\sum_M\vp_{0M}'^2}-\frac{4\pi G}{\H}\nabla^{-2}\sum_K(X_K\dvpK1+\vp_{0K}'\dvpK1')_,{}^l\Bigg]\,.
\eea

\subsection{The curvature perturbation, $\zeta$, in terms of the scalar fields}
\label{zeta_phi}

We now have the tools at hand to express the gauge-invariant curvature
perturbation on uniform density hypersurfaces in terms of the scalar
field fluctuations. Substituting Eq.~(\ref{eq:drho1}) into
Eq.~(\ref{defzeta}), and using Eq.~(\ref{0iEin1}) to relate the metric
perturbation $\phi_1$ to the scalar fields as
\be 
\phi_1=\frac{4\pi G}{\H}\sum_K \vp_{0K}'\dvpK1\,,
\ee
we obtain
\be
\zeta_1
=\frac{1}{a^2}
\left[\frac{\sum_K\left(\vp_{0K}'{\dvpK1}'+a^2 U_{,\vp_K}\dvpK1\right)}
{3\sum_L{\vp_{0L}'}^2}
-\frac{\H\sum_K\vp_{0K}'{\dvpK1}}{\sum_L{\vp_{0L}'}^2+2a^2U_0}\right]\,.
\ee

At second order, the expression is again more complicated, but we
follow the same procedure, by substituting for the metric
perturbations from Sections \ref{sec:field1} and \ref{sec:field2} and
the energy density perturbations in Eqs.~(\ref{eq:drho1}) and
(\ref{eq:drho2}) into Eq.~(\ref{zeta2scal}). We also use the Klein
Gordon equations in Section \ref{sec:KG} to substitute for
$\vp_{0K}''$ and $\dvpK1''$. This results in the following expression
for $\zeta_2$ in terms of only scalar field perturbations
\bea
\fl
\zeta_2=\frac{1}{3\sum_N\vp_{0N}'^2
}\sum_K\Bigg[\vp_{0K}'\dvpK2'-4\dvpK1'Y_K
+2Y_K^2-\frac{4\pi G}{\H}\vp_{0K}'\sum_M\vp_{0M}'\dvpM2
\nonumber \\
\fl\quad
-\frac{4\pi G}{\H^2}\vp_{0K}'\nabla^{-2}\sum_M\Big(Y_{K,kl}\nabla^{-2}(X_M\dvpM1+\vp_{0M}'\dvpM1')_{,}{}^{kl}-\nabla^2Y_K(X_M\dvpM1+\vp_{0M}'\dvpM1')\Big)
\nonumber\\
\fl\quad
-\frac{8\pi G}{\H}\vp_{0K}'^2\sum_M\nabla^{-2}\Big(\dvpM1'\nabla^2\dvpM1
+\dvpM1{}_{,l}\dvpM1'{}_{,}{}^l\Big)
+\dvpK1'^2+a^2\delta U_2+\dvpK1{}_{,l}\dvpK1{}_{,}{}^l\Bigg]
\nonumber\\
\fl\quad
-\frac{2}{3\sum_N\vp_{0N}'^2}\sum_{M}\vp_{0M}'\dvpM{}_{,l}
\Bigg[\frac{\sum_K\vp_{0K}'\dvpK1{}_{,l}}{\sum_L\vp_{0L}'^2}
-\frac{4\pi G}{\H}\nabla^{-2}\sum_K\Big(X_K\dvpK1+\vp_{0K}'\dvpK1'\Big)_{,}{}^{l}\Bigg]
\nonumber\\
\fl\quad
+\frac{2}{9\H(\sum_P\vp_{0P}'^2)^2}\sum_{K,M,N}\Bigg[\nabla^2\dvpK1\vp_{0K}'-6\H\vp_{0K}'(\dvpK1'-Y_K)
-\frac{8\pi G}{\H}\vp_{0K}'\Big(X_M\vp_{0K}'+a^2U_{,K}\vp_{0M}'\Big)\dvpM1
\nonumber\\
\fl\quad
+2a^2U_{,M}Y_M+\frac{4\pi G}{\H}\Big(a^2\vp_{0K}'^2\delta U_1+2a^2UY_K\vp_{0K}'
-\sum_L\vp_{0K}'^2\vp_{0L}'\dvpL1'\Big)\Bigg]
\Big(\vp_{0N}'(\dvpN1'-Y_N)+a^2\delta U_1\Big)
\nonumber\\
\fl\quad
+\frac{(5+3c_{\rm s}^2)}{9(\sum_P\vp_{0P}'^2)^2}\sum_{K,L}
\Bigg[\vp_{0K}'\vp_{0L}'(\dvpK1'-Y_K)(\dvpL1'-Y_L)+a^4\delta U_1^2
+2a^2\delta U_1\vp_{0K}'(\dvpK1'-Y_K)\Bigg]
\nonumber\\
\fl\quad
-\frac{4\pi G}{3\H^2\sum_P\vp_{0P}'^2}
\sum_{K,L}\Big(\vp_{0K}'\dvpK1{}_{,l}'
-\vp_{0K}'Y_{K,l}+a^2\delta U_{1,l}\Big)\nabla^{-2}(X_L\dvpL1+\vp_{0L}'\dvpL1')_,{}^l
\nonumber\\
\fl\quad
-\sum_{K,L}\frac{\vp_{0K}'\vp_{0L}'}{18\H^2\sum_P\vp_{0P}'^2}
\nabla^{-2}\Bigg\{\nabla^2\Bigg(\dvpK1'-Y_K+\frac{a^2\delta U}{\vp_{0K}'}\Bigg)\nabla^2\Bigg(\dvpL1'-Y_L+\frac{a^2\delta U}{\vp_{0L}'}\Bigg)\nonumber\\
\fl\qquad
-\Bigg(\dvpK1'-Y_K+\frac{a^2\delta U}{\vp_{0K}'}\Bigg)_{,ik}\Bigg(\dvpL1'-Y_L+\frac{a^2\delta U}{\vp_{0L}'}\Bigg)_,\Bigg.^{ik}\nonumber\\
\fl\qquad
\label{eq:zeta2full}
-{24\pi Ga}\Bigg[\Bigg(\dvpK1'-Y_K+\frac{a^2\delta U}{\vp_{0K}'}\Bigg)_,\Bigg.^{(i}\nabla^{-2}\Bigg(\frac{X_L\dvpL1}{\vp_{0L}'}+\dvpL1'\Bigg)_,\Bigg.^{j)}\Bigg]_{,ij}\Bigg\}
\eea
where we have introduced, for ease of presentation, the quantity
\be 
Y_K=\vp_{0K}'\frac{4\pi G}{\H}\sum_L\vp_{0L}'\dvpL1\,,
\ee
and the adiabatic sound speed is given by (see, for example, Refs.~\cite{nonad, Huston:2011fr})
\be 
c_{\rm s}^2=1+\frac{2}{3\H}\frac{\sum_K U_{,\vp_K}\vp_{0K}'}{\sum_L\vp_{0L}'^2}\,.
\ee

Thus, we have achieved our goal of obtaining the expression for $\zeta_2$ solely in terms of scalar field variables. While this may appear cumbersome, we should stress that this is valid for multiple scalar fields and, for the ease of numerical computation, we have reduced the expression to include only scalar field fluctuations and their first time derivative. Although in this paper we have presented our results in real space, this can be readily translated into Fourier space using, for example, the methods described in Ref.~\cite{Malik:2006ir}. 

\section{Single field, slow roll approximation}
\label{sec:results}

In this section, we will assume a single field. The slow roll approximation then allows us to relate the scalar field's potential and its derivative to the Hubble parameter and background field derivative through
\be 
2\H\vp_0'\simeq-a^2U_{,\vp}\,,
\ee
and
\be 
\H^2\simeq\frac{8\pi G}{3}a^2 U\,,
\ee
while $c_{\rm s}^2=-1$ and $X=\H\vp_0'$. Using these relationships, we can simplify Eq.~(\ref{eq:zeta2full}) to give
\bea
\fl
\zeta_2=\frac{1}{3\vp_0'}\Bigg[\dvp2'-2\H\dvp2-
\frac{4\pi G}{\H}\vp_0'^2\dvp2\Bigg]
+\frac{1}{3\vp_0'^2}\Bigg(\dvp1'^2+a^2U_{,\vp\vp}\dvp1^2
+\dvp1_,{}^l\dvp1'_{,l}\Bigg)
\nonumber\\
\fl\quad
+\frac{4}{9\vp_0'^2}\Bigg\{\Big(\dvp1'-\frac{4\pi G}{\H}\vp_0'^2\dvp1\Big)^2+9\H^2\dvp1^2-6\H\dvp1\Big(\dvp1-\frac{4\pi G}{\H}\vp_0'^2\dvp1\Big)\Bigg\}\nonumber\\
\fl
\quad
+\frac{4\pi G}{3\H}\Bigg[\frac{8\pi G}{\H}\vp_0'^2\dvp1^2
-4\dvp1\dvp1'
-\frac{4\pi G}{\H^2}\vp_0'\nabla^{-2}
\Big(\dvp1_{,kl}\nabla^{-2}\dvp1'_,{}^{kl}
-\nabla^2\dvp1\dvp1'\Big)
\nonumber\\
\fl\qquad
-2\nabla^{-2}\Big(\dvp1'\nabla^2\dvp1+\dvp1_{,l}\dvp1_{,}{}^l\Big)
-\frac{1}{\H\vp_0'}\nabla^{-2}\dvp1_{,l}'\Big(\dvp1_{,l}'-\frac{4\pi G}{\H}\vp_0'^2\dvp1_{,l}-3\H\dvp1_{,l}\Big)
\Bigg]
\nonumber\\
\fl\quad
+\frac{2}{9\H\vp_0'^2}\Bigg[\dvp1'-\frac{4\pi G}{\H}\vp_0'^2\dvp1-3\H\dvp1\Bigg]
\Bigg[\nabla^2\dvp1-6\H\dvp1'+24\pi G\vp_0'^2\dvp1-\frac{4\pi G}{\H}\vp_0'^2\dvp1'\Bigg]
\nonumber\\
\fl\quad
-\frac{2}{3\vp_0'}\dvp1_{,l}\Bigg[\frac{\dvp1_{,l}}{\vp_0'}
-\frac{4\pi G}{\H}\vp_0'\nabla^{-2}\dvp1_,{}^l{}'\Bigg]
-\frac{1}{18\H^2}\nabla^{-2}\Bigg\{\Bigg[\nabla^2\Big(\dvp1'-\frac{4\pi G}{\H}\vp_0'^2\dvp1-3\H\dvp1\Big)\Bigg]^2
\nonumber\\
\fl\qquad
-\Big(\dvp1'-\frac{4\pi G}{\H}\vp_0'^2\dvp1-3\H\dvp1\Big)_{,ij}
\Big(\dvp1'-\frac{4\pi G}{\H}\vp_0'^2\dvp1-3\H\dvp1\Big)_,\Big.^{ij}
\nonumber\\
\fl\qquad
-24\pi G\varphi_0'{}^2\Bigg[\Big(\dvp1'-\frac{4\pi G}{\H}\vp_0'^2\dvp1-3\H\dvp1\Big)_,{}\Big.^{(i}
\nabla^{-2}\dvp1'_,{}^{j)}\Bigg]_{,ij}
\Bigg\}\,.
\eea
Taking the large scale limit of this expression, where $\nabla^2\to 0$, gives
\bea
\fl
\zeta_2=\frac{1}{3\vp_0'}\Bigg[\dvp2'-2\H\dvp2-
\frac{4\pi G}{\H}\vp_0'^2\dvp2\Bigg]
+\frac{1}{3\vp_0'^2}\Bigg(\dvp1'^2+a^2U_{,\vp\vp}\dvp1^2
\Bigg)+\frac{8\pi G}{3\H}\nabla^{-2}\dvp1_,{}^l{}'\dvp1_{,l}
\nonumber\\
\fl\quad
+\frac{4}{9\vp_0'^2}\Bigg\{\Big(\dvp1'-\frac{4\pi G}{\H}\vp_0'^2\dvp1\Big)^2+9\H^2\dvp1^2-6\H\dvp1\Big(\dvp1-\frac{4\pi G}{\H}\vp_0'^2\dvp1\Big)\Bigg\}\nonumber\\
\fl
\quad
+\frac{4\pi G}{3\H}\Bigg[\frac{8\pi G}{\H}\vp_0'^2\dvp1^2
-4\dvp1\dvp1'
-\frac{4\pi G}{\H^2}\vp_0'\nabla^{-2}
\Big(\dvp1_{,kl}\nabla^{-2}\dvp1'_,{}^{kl}
-\nabla^2\dvp1\dvp1'\Big)
\nonumber\\
\fl\qquad
-2\nabla^{-2}\Big(\dvp1'\nabla^2\dvp1+\dvp1_{,l}\dvp1_{,}{}^l\Big)
-\frac{1}{\H\vp_0'}\nabla^{-2}\dvp1_{,l}'\Big(\dvp1_{,l}'-\frac{4\pi G}{\H}\vp_0'^2\dvp1_{,l}-3\H\dvp1_{,l}\Big)
\Bigg]
\nonumber\\
\label{eq:singlelargescale}
\fl\quad
+\frac{2}{9\H\vp_0'^2}\Bigg[\dvp1'-\frac{4\pi G}{\H}\vp_0'^2\dvp1-3\H\dvp1\Bigg]
\Bigg[24\pi G\vp_0'^2\dvp1-6\H\dvp1'-\frac{4\pi G}{\H}\vp_0'^2\dvp1'\Bigg]
\,.
\eea

\section{Discussion}
\label{sec:diss}

In this article, we have focused on the curvature perturbation in the uniform density gauge, $\zeta$, at second order in perturbation theory. We have presented, for the first time, the expression for $\zeta_2$ in terms of scalar fields, computed using full cosmological perturbation theory. Using the relevant Einstein field equations, we have replaced all metric perturbations, and so the resultant expression contains only the scalar field, its fluctuations and single time derivatives. 

This expression will likely be useful for future numerical computations of inflationary observables, such as the bispectrum; we have presented the final expression in such a way to make the numerical implementation as simple as possible. Additionally, we have presented the case for a single, slowly-rolling scalar field, and have then taken the large scale limit of this expression. We can see the benefit of using full perturbation theory from this expression, Eq.~(\ref{eq:singlelargescale}), where we obtain terms containing inverse Laplacians, which would not be present in the corresponding expression from the $\delta N$ approach. These terms can be removed, on assuming that the metric perturbation $B$ is decaying at both first and second order, allowing one to use the momentum equation to replace the inverse Laplacian terms. However, this is only the case on scales much larger than the horizon. We present the more general equation, valid on all scales, in this work.

While this project was in its final stages we became aware of a study obtaining similar results, where comparison is possible \cite{Dias:2014msa}.

\ack{
AJC is funded by the Sir Norman Lockyer Fellowship of the Royal Astronomical Society, KAM is supported, in part, by STFC grant ST/J001546/1 and EN is funded by an STFC studentship. {\sc Cadabra} \cite{Peeters:2007wn, Peeters:2006kp} was used in the derivation of some equations. The authors are grateful to Laila Alabidi and Ian Huston for early ideas on this topic, to David Mulryne for useful discussions, and to Pedro Carrilho for comments on an early draft of the manuscript. AJC thanks the Astronomy Unit at QMUL for hospitality at various stages of the project.}

\appendix

\section{Second order Klein-Gordon Equation}

In this section, for completeness, we present the Klein-Gordon equation for the second order scalar field, in a close form, from Ref.~\cite{Malik:2006ir}.

\bea
\fl
\label{flatKG2real}
\dvpI2''+2\H\dvpI2'-\nabla^2\dvpI2
+a^2\sum_K\left[
U_{,\vp_K\vp_I}+\frac{8 \pi G}{\H}\left(
\vp_{0I}'U_{,\vp_K}+\vp_{0K}'U_{,\vp_I}
+\vp_{0K}'\vp_{0I}'\frac{8 \pi G}{\H}U_0
\right)
\right]\dvpK2 \nonumber\\
\fl
+\frac{16\pi G}{\H}\Bigg[\
 \dvpI1' \sum_K X_K\dvpK1
+\sum_K\vp_{0K}'\dvpK1 \sum_K a^2 U_{,\vp_I\vp_K}\dvpK1
\Bigg] \nonumber\\
\fl
+\left(\frac{8\pi G}{\H}\right)^2 \sum_K\vp_{0K}'\dvpK1
\Bigg[\
a^2 U_{,\vp_I}\sum_K\vp_{0K}'\dvpK1
+\vp_{0I}'
\sum_K\left(a^2 U_{,\vp_K}+X_K\right)\dvpK1
\Bigg]\nonumber\\
\fl
-2\left(\frac{4\pi G}{\H}\right)^2\frac{\vp_{0I}'}{\H}
\sum_K X_K\dvpK1 \sum_K \left( X_K\dvpK1
+ \vp_{0K}'\dvpK1'\right)
+\frac{4\pi G}{\H}\vp_{0I}'\sum_K{\dvpK1'}^2
\nonumber\\
\fl
+a^2\sum_{K,L} \left[
U_{,\vp_I\vp_K\vp_L} 
+ \frac{8\pi G}{\H}\vp_{0I}' U_{,\vp_K\vp_L}
\right]\dvpK1\dvpL1
+F\left(\dvpK1',\dvpK1\right)=0\,,
\eea
where $F\left(\dvpK1',\dvpK1\right)$ contains gradients and inverse
gradients quadratic in the field fluctuations and is defined as
\bea
\label{Fdvk1}
\fl\hspace{-5mm}
F\left(\dvpK1',\dvpK1\right)
%
=
\left(\frac{8\pi G}{\H}\right)^2
\delta\vp_{1I,l}'\nabla^{-2}\sum_K\left(
X_K\dvpK1+\vp_{0K}'\dvpK1'\right)_{,}^{~l}
-\frac{16\pi G}{\H}\nabla^2\dvpI1\sum_K\vp_{0K}'\dvpK1
\nonumber\\
\hspace{-5mm}
+2\frac{X_I}{\H}\left(\frac{4\pi G}{\H}\right)^2 
\nabla^{-2}
\Bigg[
\sum_K\vp_{0K}'\delta\vp_{1K,lm}
\nabla^{-2}\sum_K\left(X_K\dvpK1+\vp_{0K}'\dvpK1'\right)_{,}^{~lm}
\nonumber\\
\qquad
-\sum_K\left(X_K\dvpK1+\vp_{0K}'\dvpK1'\right)
\nabla^{2}\sum_K\vp_{0K}'\dvpK1
\Bigg]
\nonumber\\
\hspace{-5mm}
+\frac{4\pi G}{\H}
\Bigg[\vp_{0I}'\sum_K \delta\vp_{1K,l}\delta\vp_{1K,}^{~~~~l}
+4X_I\nabla^{-2}\sum_K\left(
\dvpK1'\nabla^2\dvpK1+\dvpK1'_{,l}\dvpK1_{,}^{~l}\right)
\Bigg]
\nonumber\\
\hspace{-5mm}
+\left(\frac{4\pi G}{\H}\right)^2
\frac{\vp_{0I}'}{\H}\Bigg[
\nabla^{-2}\sum_K\left( X_K\dvpK1+\vp_{0K}'\dvpK1'\right)_{,lm}
\nabla^{-2}\sum_K\left(X_K\dvpK1+\vp_{0K}'\dvpK1'\right)_{,}^{~lm}
\nonumber\\
\qquad
-\sum_K\vp_{0K}'\delta\vp_{1K,l}
\sum_K\vp_{0K}'\delta\vp_{1K,}^{~~~~l}
\Bigg]\nonumber\\
\fl
\hspace{-5mm}
-\frac{\vp_{0I}'}{\H} \nabla^{-2}
\Bigg\{
8\pi G\sum_K\left(\dvpK1_{,l}\nabla^{2}\dvpK1_{,}^{~l}
+\nabla^{2}\dvpK1\nabla^{2}\dvpK1
+\dvpK1'\nabla^{2}\dvpK1'+\dvpK1'_{,l}\dvpK1_{,}^{\prime~l}\right)
\nonumber\\
\fl
-\left(\frac{4\pi G}{\H}\right)^2
\Bigg[
2\nabla^{-2}\sum_K\left(X_K\dvpK1+\vp_{0K}'\dvpK1'\right)_{,~j}^{~i}
\sum_K X_K\dvpK1
+\sum_K\vp_{0K}'\dvpK1_{,}^{~i}\sum_K\vp_{0K}'\dvpK1_{,j}
\Bigg]_{,i}^{~j}
\Bigg\}\,.
\eea

\section*{References}
\bibliographystyle{unsrt.bst}
\bibliography{zeta_2_ajc_final.bbl}

\begin{thebibliography}{10}

\bibitem{COBE}
C.~L. Bennett et~al.
\newblock {Cosmic temperature fluctuations from two years of COBE differential
  microwave radiometers observations}.
\newblock {\em Astrophys. J.}, 436:423--442, 1994.

\bibitem{WMAP7}
E.~Komatsu et~al.
\newblock {Seven-Year Wilkinson Microwave Anisotropy Probe (WMAP) Observations:
  Cosmological Interpretation}.
\newblock 2010.

\bibitem{Ade:2013uln}
P.A.R. Ade et~al.
\newblock {Planck 2013 results. XXII. Constraints on inflation}.
\newblock 2013.

\bibitem{Ade:2014xna}
P.A.R. Ade et~al.
\newblock {BICEP2 I: Detection Of B-mode Polarization at Degree Angular
  Scales}.
\newblock 2014.

\bibitem{Ade:2015tva}
P.A.R. Ade et~al.
\newblock {A Joint Analysis of BICEP2/Keck Array and Planck Data}.
\newblock {\em Phys.Rev.Lett.}, 2015.

\bibitem{Baumann:2009ds}
Daniel Baumann.
\newblock {TASI Lectures on Inflation}.
\newblock 2009.

\bibitem{Bardeen:1980kt}
James~M. Bardeen.
\newblock {Gauge Invariant Cosmological Perturbations}.
\newblock {\em Phys. Rev.}, D22:1882--1905, 1980.

\bibitem{ks}
Hideo Kodama and Misao Sasaki.
\newblock {Cosmological Perturbation Theory}.
\newblock {\em Prog. Theor. Phys. Suppl.}, 78:1--166, 1984.

\bibitem{mfb}
Viatcheslav~F. Mukhanov, H.~A. Feldman, and Robert~H. Brandenberger.
\newblock {Theory of cosmological perturbations. Part 1. Classical
  perturbations. Part 2. Quantum theory of perturbations. Part 3. Extensions}.
\newblock {\em Phys. Rept.}, 215:203--333, 1992.

\bibitem{Starobinsky:1986fxa}
Alexei~A. Starobinsky.
\newblock {Multicomponent de Sitter (Inflationary) Stages and the Generation of
  Perturbations}.
\newblock {\em JETP Lett.}, 42:152--155, 1985.

\bibitem{Sasaki:1995aw}
Misao Sasaki and Ewan~D. Stewart.
\newblock {A General analytic formula for the spectral index of the density
  perturbations produced during inflation}.
\newblock {\em Prog.Theor.Phys.}, 95:71--78, 1996.

\bibitem{Sasaki:1998ug}
Misao Sasaki and Takahiro Tanaka.
\newblock {Superhorizon scale dynamics of multiscalar inflation}.
\newblock {\em Prog.Theor.Phys.}, 99:763--782, 1998.

\bibitem{Wands2000}
David Wands, Karim~A. Malik, David~H. Lyth, and Andrew~R. Liddle.
\newblock {A new approach to the evolution of cosmological perturbations on
  large scales}.
\newblock {\em Phys. Rev.}, D62:043527, 2000.

\bibitem{Rigopoulos:2003ak}
G.I. Rigopoulos and E.P.S. Shellard.
\newblock {The separate universe approach and the evolution of nonlinear
  superhorizon cosmological perturbations}.
\newblock {\em Phys.Rev.}, D68:123518, 2003.

\bibitem{Lyth:2004gb}
David~H. Lyth, Karim~A. Malik, and Misao Sasaki.
\newblock {A general proof of the conservation of the curvature perturbation}.
\newblock {\em JCAP}, 0505:004, 2005.

\bibitem{Malik2004}
Karim~A Malik and David Wands.
\newblock {Evolution of second-order cosmological perturbations}.
\newblock {\em Class. Quant. Grav.}, 21:L65--L72, 2004.

\bibitem{Lyth:2005du}
David~H. Lyth and Yeinzon Rodriguez.
\newblock {Non-Gaussianity from the second-order cosmological perturbation}.
\newblock {\em Phys.Rev.}, D71:123508, 2005.

\bibitem{Malik:2005cy}
Karim~A. Malik.
\newblock {Gauge-invariant perturbations at second order: multiple scalar
  fields on large scales}.
\newblock {\em JCAP}, 0511:005, 2005.

\bibitem{Langlois:2005qp}
David Langlois and Filippo Vernizzi.
\newblock {Conserved non-linear quantities in cosmology}.
\newblock {\em Phys.Rev.}, D72:103501, 2005.

\bibitem{Nakamura:2010yg}
Kouji Nakamura.
\newblock {Second-order Gauge-Invariant Cosmological Perturbation Theory:
  Current Status}.
\newblock {\em Adv.Astron.}, 2010:576273, 2010.

\bibitem{Tzavara:2011hn}
Eleftheria Tzavara and Bartjan van Tent.
\newblock {Gauge-invariant perturbations at second order in two-field
  inflation}.
\newblock {\em JCAP}, 1208:023, 2012.

\bibitem{MW2008}
Karim~A. Malik and David Wands.
\newblock {Cosmological perturbations}.
\newblock {\em Phys. Rept.}, 475:1--51, 2009.

\bibitem{thesis}
Adam~J. Christopherson.
\newblock {\em {Applications of Cosmological Perturbation Theory}}.
\newblock PhD thesis, University of London, 2011.
\newblock * Temporary entry *.

\bibitem{Malik:2006ir}
Karim~A. Malik.
\newblock {A not so short note on the Klein-Gordon equation at second order}.
\newblock {\em JCAP}, 0703:004, 2007.

\bibitem{Lyth:1984gv}
D.H. Lyth.
\newblock {Large Scale Energy Density Perturbations and Inflation}.
\newblock {\em Phys.Rev.}, D31:1792--1798, 1985.

\bibitem{Bardeen:1983qw}
James~M. Bardeen, Paul~J. Steinhardt, and Michael~S. Turner.
\newblock {Spontaneous Creation of Almost Scale - Free Density Perturbations in
  an Inflationary Universe}.
\newblock {\em Phys. Rev.}, D28:679, 1983.

\bibitem{Salopek:1990jq}
D.~S. Salopek and J.~R. Bond.
\newblock {Nonlinear evolution of long wavelength metric fluctuations in
  inflationary models}.
\newblock {\em Phys. Rev.}, D42:3936--3962, 1990.

\bibitem{nonad}
Adam~J. Christopherson and Karim~A. Malik.
\newblock {The non-adiabatic pressure in general scalar field systems}.
\newblock {\em Phys. Lett.}, B675:159--163, 2009.

\bibitem{Huston:2011fr}
Ian Huston and Adam~J. Christopherson.
\newblock {Calculating Non-adiabatic Pressure Perturbations during Multi-field
  Inflation}.
\newblock {\em Phys.Rev.}, D85:063507, 2012.

\bibitem{Dias:2014msa}
Mafalda Dias, Joseph Elliston, Jonathan Frazer, David Mulryne, and David Seery.
\newblock {The curvature perturbation at second order}.
\newblock 2014.

\bibitem{Peeters:2007wn}
Kasper Peeters.
\newblock {Introducing Cadabra: A Symbolic computer algebra system for field
  theory problems}.
\newblock 2007.

\bibitem{Peeters:2006kp}
Kasper Peeters.
\newblock {A Field-theory motivated approach to symbolic computer algebra}.
\newblock {\em Comput.Phys.Commun.}, 176:550--558, 2007.

\end{thebibliography}

\end{document}